\documentclass[letterpaper,twocolumn,english,superscriptaddress,nofootinbib]{revtex4-1}

\usepackage[]{fontenc}
\usepackage[latin9]{inputenc}
\usepackage{textcomp}
\usepackage{amsmath}
\usepackage{graphicx}
\usepackage{amssymb}
\usepackage{color}
\usepackage{soul}

\makeatletter

\pdfpageheight\paperheight
\pdfpagewidth\paperwidth

\newcommand{\lyxmathsym}[1]{\ifmmode\begingroup\def\b@ld{bold}
  \text{\ifx\math@version\b@ld\bfseries\fi#1}\endgroup\else#1\fi}

\@ifundefined{textcolor}{}
{%
 \definecolor{BLACK}{gray}{0}
 \definecolor{WHITE}{gray}{1}
 \definecolor{RED}{rgb}{1,0,0}
 \definecolor{GREEN}{rgb}{0,1,0}
 \definecolor{BLUE}{rgb}{0,0,1}
 \definecolor{CYAN}{cmyk}{1,0,0,0}
 \definecolor{MAGENTA}{cmyk}{0,1,0,0}
 \definecolor{YELLOW}{cmyk}{0,0,1,0}
 }
 \newcommand{\ya}{$\mathrm{YbAl_3}$}
\newcommand{\la}{$\mathrm{LuAl_3}$}

\newcommand{\fo}{$\mathrm{\textit{f}^{13}}$}
\newcommand{\ft}{$\mathrm{\textit{f}^{12}}$}

\makeatother

\usepackage{babel}

\begin{document}

\title{Lifshitz Transition from Valence Fluctuations in YbAl$_{3}$} 

\author{Shouvik Chatterjee}
\altaffiliation[Current address: ]{Department of Electrical $\&$ Computer Engineering, University of California, Santa Barbara, CA 93106, USA}
\affiliation{Laboratory of Atomic and Solid State Physics, Department of Physics, Cornell University, Ithaca, New York 14853, USA}

\author{Jacob P. Ruf}
\affiliation{Laboratory of Atomic and Solid State Physics, Department of Physics, Cornell University, Ithaca, New York 14853, USA}

\author{Haofei I. Wei}
\affiliation{Laboratory of Atomic and Solid State Physics, Department of Physics, Cornell University, Ithaca, New York 14853, USA}

\author{Kenneth D. Finkelstein}
\affiliation{Cornell High Energy Synchrotron Source, Wilson Laboratory, Cornell University, Ithaca, New York 14853, USA}

\author{Darrell G. Schlom}

\affiliation{Department of Materials Science and Engineering, Cornell University, Ithaca, New York 14853, USA}
\affiliation{Kavli Institute at Cornell for Nanoscale Science, Ithaca, New York 14853, USA}

\author{Kyle M. Shen}

\email[email: ]{kmshen@cornell.edu}

\affiliation{Laboratory of Atomic and Solid State Physics, Department of Physics, Cornell University, Ithaca, New York 14853, USA}
\affiliation{Kavli Institute at Cornell for Nanoscale Science, Ithaca, New York 14853, USA}

\maketitle


\textbf{In Kondo lattice systems with mixed valence, such as YbAl$_{3}$, interactions between localized electrons in a partially filled \textit{f} shell and delocalized conduction electrons can lead to fluctuations between two different valence configurations with changing temperature or pressure. The impact of this change on the momentum-space electronic structure and Fermi surface topology is essential for understanding their emergent properties, but has remained enigmatic due to a lack of appropriate experimental probes. Here by employing a combination of molecular beam epitaxy (MBE) and \textit{in situ} angle-resolved photoemission spectroscopy (ARPES) we show that valence fluctuations can lead to dramatic changes in the Fermi surface topology, even resulting in a Lifshitz transition. As the temperature is lowered, a small electron pocket in YbAl$_{3}$ becomes completely unoccupied while the low-energy ytterbium (Yb) 4\textit{f} states become increasingly itinerant, acquiring additional spectral weight, longer lifetimes, and well-defined dispersions. Our work presents the first unified picture of how local valence fluctuations connect to momentum space concepts including band filling and Fermi surface topology in the longstanding problem of mixed-valence systems.}


\section{Introduction}

Kondo lattice systems host a wide variety of quantum states such as antiferromagnetism\cite{Schroeder:00}, heavy Fermi liquids\cite{Andres:75}, hidden order\cite{Mydosh:11}, and unconventional superconductivity\cite{Curro:05}, which can often be controlled by modest perturbations using magnetic field or pressure, thereby providing  access to quantum phase transitions\cite{Si:01,Gegenwart:08,Si:10}. These states generally emerge from a complex many-body state that is formed by enhanced Kondo coupling between the local rare-earth moments and the band-like conduction electrons at low temperatures. In mixed valence systems\cite{Varma:76, Lawrence:80, Parks:77}, this coupling also results in a change of the rare-earth valence, which can be determined by core-level spectroscopies that probe the local chemical environment (\textit{r} - space)\cite{Tjeng:93,Moreschini:07,Kummer:11}, but the implications for the momentum-space (\textit{k} - space) electronic structure remain poorly understood. To gain insight into the emergent properties of these systems, it is crucial to understand how delocalized carriers and the low-energy momentum-space electronic structure emerge from these local interactions. 

Here, we choose \ya\/ as a simple prototypical mixed valence system with two nearly degenerate ytterbium (Yb) valence configurations, Yb$^{2+}$ (4\textit{f}$^{14}$) and Yb$^{3+}$(4\textit{f}$^{13}$). The average Yb valence, $\nu_{f}$, decreases with temperature, changing by approximately $\Delta \nu_{f}$ = -0.05 from 300 K to below T$^{*}$ $\approx$ 34 - 40 K\cite{Tjeng:93, Moreschini:07, Suga:05, Bauer:04, Lawrence:94}, when it becomes a heavy Fermi liquid, attributed to the enhanced Kondo screening at low temperatures\cite{Cornelius:02}. We selected \ya\/ due to its relatively large change in valence as well as its large energy scales, with a reported single ion Kondo temperature T$_{K}$ $\approx$ 670 K\cite{Cornelius:02,Ebihara:03}, which should make these changes observable in momentum space. The lack of a well-defined, pristine surface in cleaved \ya\/ single crystals\cite{Wahl:11}, however, has previously prevented momentum-resolved measurements of its electronic structure. We have circumvented this problem by synthesizing for the first time, epitaxial thin films of \ya\/ and its conventional metal analogue \la\/ by MBE\cite{Chatterjee:16} and have combined it with \textit{in situ} ARPES to directly measure their electronic structure as a function of temperature.

Our measurements reveal a strong temperature dependent change in both the real and momentum-space electronic structure of \ya\/. The local Yb valence decreases as the temperature is lowered, accompanied by a large shift in the chemical potential which leads to a Lifshitz transition of a small electron pocket at $\Gamma$, along with the emergence of renormalized heavy quasiparticles near the Fermi energy (E$_{F}$). We establish, for the first time, a direct one-to-one correspondence between these observed changes, which we believe to be generic to all mixed-valence systems.

\section{Results and Discussion}

\subsection{Synthesis and Electronic Structure}

Both \ya\/ and \la\/ crystallize in a cubic \textit {Pm$\overline{3}$m} structure where Yb or Lu atoms occupy the vertices of the unit cell while Al atoms occupy the face centers, as illustrated in the inset of Fig. 1b. \la\/ has fully occupied 4\textit{f} orbitals with zero net moment and a lattice constant (4.19 \AA) closely matched to \ya\/ (4.20 \AA). Thus, \la\/ serves as an ideal reference compound to understand the light, Al-derived band-like conduction electron states, which are also common to \ya\/. Epitaxial thin films of both \la\/ and \ya\/ with (001) out-of-plane orientation were synthesized by co-evaporation on MgO (001) substrates (4.21 \AA\/) at temperatures of 200 - 350$^{\circ}$C and a chamber base pressure below 2 $\times 10^{-9}$ torr. For all films, a 1.2 nm thick aluminum (4.05 \AA\/) buffer layer was deposited at 500$^{\circ}$C, which allowed the growth of continuous, smooth films of \la\//\ \ya\/ on top. In these studies, we investigated a 30 nm thick \la\/ film and a 20 nm thick \ya\/ film (the \ya\/ was synthesized on top of a 20 nm thick \la\/ buffer layer on top of the Al buffer, which improved the quality of the \ya\/ layers). All films were sufficiently thick so that any photoemission intensity from the buffer layers or substrate and thickness dependent finite size effects can be ignored. Additional details about the synthesis can be found in the Methods section as well in Ref. \onlinecite{Chatterjee:16}.

In Fig. 1, we show Fermi surface maps and the electronic structure from \la\/ and \ya\/ thin films from the Fermi energy (E$_{F}$) to a binding energy of 10.5 eV. For \la\/, only the Lu 4\fo\/ final states are observed, with the $J = 7/2$ and $5/2$ core levels at binding energies of 6.7 and 8.2 eV, respectively. Highly dispersive Al-derived bands can be observed in both \la\/ (Fig. 1c) and \ya\/ (Fig. 1f), which extend from about 6 eV binding energy to near E$_{F}$. By matching the experimentally determined dispersion of these bands as well the Fermi surface contours measured with both He I$\alpha$ (21.2 eV) and He II$\alpha$ (40.8 eV) photons to density functional theory (DFT) calculations, we are able to determine our $k_{z}$ values for both \ya\/ and \la. Due to the lack of strong correlations in \la\/ (its 4$f$ shell is entirely filled), density functional theory (DFT) calculations should accurately describe its electronic structure and indeed we find good agreement between both the DFT-calculated band dispersions as well as Fermi surface contours to the experimentally determined dispersions and Fermi surface from ARPES, assuming an inner potential of $V_{0}$ = 18 eV (Fig. 1a,c and Fig. S1). We observe broadly dispersive, primarily Al-derived bands in \ya\/ (Fig. 1f) analogous to those observed in \la\/, and also found excellent correspondence between the measured electronic structure in \la\/ and \ya\/ over the entire Brillouin zone (Fig. S3) indicating that we are probing a similar $k_{z}$ as in \la\/, as one might expect given their highly similar electronic and crystal structures. Using the value of $V_{0}$, we determine that for $h\nu$ = 21.2 eV, we are probing near the zone center, $\Gamma$, $k_{z}$ = 0 $\pm$ 0.1 $\pi$/c. More details about the $k_z$ determination can be found in the supporting materials\cite{Supplementary}.

A two dimensional slice at $k_{z} = \Gamma$ of the three-dimensional Fermi surface of \la\/ accesses a multiply-connected Fermi surface sheet consisting of electron-like pockets centered at (0, 0) and ($\pi$, $\pi$), consistent with our ARPES data, shown in Fig. 1a-c. On the other hand, in \ya\/, we clearly observe both the Yb 4\fo\/ and 4\ft\/ final states around 0 - 2 eV and 6 - 10.5 eV binding energy, respectively, consistent with its mixed valence character.  The near E$_{F}$ electronic structure in \ya\/ is, however, significantly modified by a shift in its chemical potential due to the differing average Lu and Yb valence and the interaction between the broad, dispersive bands and the renormalized Yb 4\textit{f} states. In the Fermi surface map of \ya\/ (Fig. 1d), large Fermi surface sheets are prominent and centered at zone edges ($\pi$, $\pi$).

\subsection{Evolution of the electron pocket at $\Gamma$}

Having discussed the basic electronic structure, we now turn towards its temperature dependence in \ya\/.  In Fig. 2a we show a series of ARPES spectra obtained along $(0, 0)$ to $(0, \pi)$ at $k_{z} \approx \Gamma$ between 255 and 21 K, which establish a clear temperature-dependent shift of the chemical potential $\Delta \mu$ with the $4f$ derived states moving closer to $E_{F}$ as the temperature is lowered, consistent with earlier angle-integrated measurements\cite{Suga:05}. The most dramatic effect of $\Delta \mu$ is on a small parabolic electron pocket centered at $\Gamma$. At 255 K, the electron pocket can be clearly observed with its band bottom at $40 \pm 5$ meV binding energy and a $k_{F}$ of $0.20 \pm 0.01$ $\pi/a$. As the temperature is lowered, the electron pocket is lifted in energy and becomes entirely unoccupied around 21 K. Since the pocket is centered at $\Gamma$, its lifting above $E_F$ would then coincide with a Lifshitz transition. To within experimental resolution, the dispersion or effective mass of the electron pocket does not change apart from a rigid shift due to $\Delta \mu$. Furthermore, while the Yb 4\fo\/ final states also shifted in energy, the Yb 4\ft\/ states did not shift appreciably with temperature indicating that the $\Delta \mu$ shift arises from an alteration in band filling due to the emergence of a Kondo screened many body state. 

We note that the small electron pocket at $\Gamma$ is not reported in previous de Haas-van Alphen (dHvA) studies of \ya\/, which can be explained by the fact that the electron pocket is only occupied at higher temperatures ($T > 20$ K), whereas dHvA measurements are conducted at low temperatures (T $\approx$ 20 mK). Nevertheless, DFT calculations suggest the presence of a quasi-spherical electron pocket at $\Gamma$, but whose size strongly depends on the binding energy of the Yb 4\textit{f} states (see Fig. S4), and thus the value of $U$ used in the calculations. Furthermore, our observation of the temperature-dependent chemical potential shift may explain the need to artificially shift the chemical potential in previous low-temperature quantum oscillation experiments of \ya\/ which compared their results to band structure calculations \cite{Ebihara:00}. The large Fermi surface sheets centered at ($\pi$, $\pi$) \cite{Supplementary} measured by ARPES would give an oscillation frequency of $\approx 1.0 \pm 0.2 \times 10^{8}$ Oe, which is comparable but somewhat larger than the largest reported  quantum oscillation frequency along the (100) direction, $6.51 \times 10^{7}$ Oe. This discrepancy might be due to the fact that the quantum oscillations are measuring a closed Fermi surface contour at a different $k_z$ than our ARPES measurements at $k_z$ = 0, which might correspond to an open contour. We also observed another Fermi surface sheet centered at the M point ($\pi$,$\pi$) with 40.8 eV photon energy which corresponds to an oscillation frequency of 3.5$\times$10$^{7}$ $\pm$ 1 $\times$ 10$^{7}$, roughly consistent with the $\beta$ pocket reported in the dHvA measurements ($4.55 \times 10^{7}$ Oe).

In Fig. 2b, we show a series of angle-integrated wide energy valence band \textit{in situ} x-ray photoemission spectra (XPS) showing a dramatic temperature-dependent change in relative intensity of the 4\fo\/ and 4\ft\/ final states. As the temperature is lowered, the relative intensity of the 4\fo\/ final states increases, while that of the 4\ft\/ final states decreases indicating a reduction of the effective Yb valence in \ya\/ at lower temperatures, which is found to be $\Delta \nu_{f} \approx 0.05$ from room temperature to below $\approx$ 45 K, in agreement with previously reported results from bulk samples\cite{Moreschini:07,Suga:05,Bauer:04,Lawrence:94,Tjeng:93}.

\subsection{Relation between real-space and momentum-space electronic structure}

In Fig. 2c, we make a quantitative comparison between the observed change in the temperature-dependent band filling and the estimated change in the Yb valence from core level spectroscopy, both in our thin films and previous measurements on \ya\/ single crystals. The change in average Yb valence in our thin films has been estimated by resonant x-ray emission spectroscopy (RXES) and XPS, details for which can be found in the Methods section and in the supporting materials\cite{Supplementary}. Assuming a spherical geometry ($\frac{4}{3}\pi k_{F}^{3}$) due to its location at $k = (0,0,0)$ and the cubic symmetry, we plot the change in Luttinger volume of the electron pocket $\Delta \nu_{Lutt}$, versus the estimated change in Yb valence $\Delta \nu_{f}$, from core level spectroscopy. Without any adjustable parameters or scaling factors, we discover a precise, one-to-one correspondence between $\Delta \nu_{Lutt}$ from the electron pocket and $\Delta \nu_{f}$ as a function of temperature. This provides the first direct microscopic evidence that in \ya\/, the Kondo screening of the $4f$ moments by the conduction electrons that results in the emergence of composite heavy fermion quasiparticles leads to a Lifshitz transition of the Fermi surface, which is also reflected in the reduction of the average Yb valence, and should be generic to other mixed valence systems. 

A qualitative model of the temperature dependent changes in both real and momentum-space is presented in Fig. 2d. As the temperature is lowered, the filling of the small electron pocket is gradually reduced as those electrons are transferred into the Kondo screening cloud at the Yb site leading to the formation of renormalized Kondo screened many-body states\cite{Choi:12} and a reduction of the effective Yb valence as measured by XPS and RXES studies. This model would explain the direct one-to-one correspondence between the measured changes in both the Yb valence and the Luttinger volume of the electron pocket as a function of temperature.

We should note that previous studies of other mixed valence systems, such as YbRh$_{2}$Si$_{2}$ have not reported temperature-dependent changes in the band structure or Fermi surface topology,\cite{Kummer:15} although this may have been because of the much larger $\Delta \nu_{f}$ in \ya\/ (0.05 versus 0.01) in the accessed temperature range of the experiments, as well as its larger energy scales ($T_{K} = 670$ K versus 25 K)\cite{Kummer:15}.


\subsection{Evolution of the Yb 4\textit{f} states}

We now discuss the evolution with temperature of the $4f$-derived heavy bands near $E_{F}$. In Fig. 3a, we show representative energy distribution curves (EDCs) at different temperatures integrated over the momentum region indicated in Fig. 3d, together with extracted changes of the $4f$ binding energy, quasiparticle weight, and scattering rate as a function of temperature (Figs. 3b,c)\cite{Supplementary}. We find a dramatic enhancement of the quasiparticle spectral weight of the $4f$ bands, consistent with previous measurements by Tjeng \emph{et al.} \cite{Tjeng:93}, coinciding with a precipitous drop in the scattering rate, which saturates around T$^{*}$ $\approx$ 37 K, the estimated coherence temperature of \ya\/\cite{Chatterjee:16} when it becomes a Fermi liquid. The enhancement of the quasiparticle spectral weight and lifetime  with decreasing temperature suggests that the screening of the $4f$ moments by the conduction electrons has nearly saturated around T$^{*}$, and that the Lifshitz transition is coincident with this dramatic change in the $4f$ spectral function. This is further highlighted by the observation of a ln($\frac{T_{0}}{T}$) (T$_{0}$ $\geq$ 255 K) scaling behavior in the integrated spectral weight, as expected from a two fluid model\cite{Choi:12,Yang_1:08,Yang_2:08,Yang:11}, until the onset of Fermi liquid behavior at T$^{*}$, when it starts to saturate. The observation of this scaling behavior up to 255 K, the highest temperature accessed in this study, suggests that the hybridization between the local 4\textit{f} moments and the conduction electrons sets in at a relatively high temperature, even though the Fermi liquid regime exists only below T$^{*}$, consistent with the slow crossover scenario predicted by slave boson mean field calculations\cite{Burdin:09,Burdin:00}. The saturation of the 4\textit{f} quasiparticle lifetime at T$^{*}$ in our ARPES measurements is also consistent with earlier transport and thermodynamic measurements, which suggested that T$^{*}$ could be related to the formation of coherence in the $4f$ states\cite{Cornelius:02,Bauer:04,Ebihara:03}, which we establish spectroscopically. The shift in binding energy of the $4f$ states is smaller compared to the $\Delta \mu$ measured from the electron-like band, with the discrepancy increasing at lower temperatures, shown in Fig. 3b, indicative of enhanced hybridization between the 4\textit{f} states and the conduction electrons at lower temperatures that pushes the electron pocket further towards lower binding energy.\\

\subsection{Dispersive crystal electric field split states}

In addition to the strong temperature dependence of the electronic structure, our measurements clearly show three distinct flat bands close to $E_{F}$ which acquire significant dispersion at certain $k$ points. Their proximity to $E_{F}$, the value of their splittings ($\approx$ 0 - 30 meV), and their narrow bandwidths are all consistent with these being crystal electric field (CEF) split states. The dispersion of the CEF states occur when the light Al-derived bands cross the flat $4f$ states near $E_{F}$, lifting the degeneracy of the CEF split branches, as shown in Fig. 3d (also see Fig. S8). The observation of three distinct bands is consistent with the bulk cubic symmetry of the Yb ions, where the Yb $J = 7/2$ manifold should split into three crystal field levels $\Gamma_{6}$, $\Gamma_{7}$, and $\Gamma_{8}$\cite{Lea:62}. While we cannot determine conclusively whether these states are representative of bulk versus surface Yb atoms, the values of their splittings (0 - 30 meV) and bandwidths ($\leq$ 25 meV) from ARPES and the fact that they extend from E$_{F}$ to a binding energy of $\approx$ 50 meV are also consistent with reports from bulk-sensitive inelastic neutron scattering which do not observe sharp CEF excitations but rather a broad continuum ($\approx 50$ meV), since those measurements would average the dispersion of the CEF states over the entire Brillouin zone \cite{Murani:94,Osborn:99,Christianson:06}.\\

\subsection{Conclusion}

Our work provides the first experimentally unified picture of how local changes of the rare earth valence impacts the momentum-space electronic structure in the prototypical mixed valence system, \ya\/. We have achieved this by combining state-of-the-art materials synthesis and advanced \textit{in situ} spectroscopy, which should be readily extendable to other Kondo lattice systems or even artificial \textit{f}-electron heterostructures. We have discovered that a Lifshitz transition of a small electron Fermi surface accompanies the change in average Yb valence, which had hitherto been unanticipated. This discovery underscores how the Kondo screening process can significantly alter $k$-space instabilities of Kondo lattice systems. 

\newpage

\section*{Methods}

\subsection{Film Growth and Characterizaion}

Single crystalline, epitaxial, atomically smooth thin films of (001) \ya\/ and \la\/ were synthesized on MgO substrates in a Veeco GEN10 MBE system with a liquid nitrogen cooled cryoshroud at a base pressure better than 2 $\times 10^{-9}$ torr. Prior to growth, MgO substrates were annealed in vacuum for 20 minutes at 800 $^{\circ}$C and a 1-2 nm thick aluminum (Al) buffer layer was deposited at 500 $^{\circ}$C. Lu/Yb and Al were co-evaporated from Langmuir effusion cells at a rate of $\approx$ 0.4 nm / minute onto a rotating substrate between 200-350$^{\circ}$ C with real-time reflection high-energy electron diffraction (RHEED) monitoring. Due to the co-evaporation growth, the surface termination was not deliberately controlled. After growth, the films were immediately transferred under ultra-high vacuum to an ARPES chamber for measurements. All ARPES data presented in this study were performed on 30 nm thick \la\/ thin films with a 1.2 nm thick Al buffer layer, or on 20 nm thick \ya\/ thin films with 20 nm thick \la\/ and 1.2 nm thick Al buffer layers. The ARPES spectra did not show any thickness dependence for \la\// \ya\/ layers that were more than 10 nm thick, the minimum thickness for this study. For further details regarding film growth and characterization see \cite{Chatterjee:16}

\subsection{\textit{In situ} ARPES and XPS}

After growth, thin film samples were immediately transferred within 5 minutes through ultra-high vacuum into an analysis chamber consisting of a VG Scienta R4000 electron analyzer, VUV5000 helium plasma discharge lamp and a dual anode x-ray source for ARPES and XPS measurements. The base pressure of the analysis chamber was better than 5$\times$10$^{-11}$ torr. ARPES measurements were performed using He I$\alpha$ (h$\nu$ = 21.2 eV) and He II$\alpha$ (h$\nu$ = 40.8 eV) photons, while Al K$\alpha$ (h$\nu$ = 1486.6 eV) photons were utilized for collecting XPS data. A polycrystalline gold reference, in electrical contact with the sample was used to determine position of the Fermi level and the energy resolution.

\subsection{DFT calculations}

DFT calculations of the band structure and Fermi surface of \la\/ / \ya\/ with were performed using full potential linearized augmented plane wave method as implemented in the Wien2k software package\cite{Wien2k}. The exchange and correlation effects were taken into account within the generalized gradient approximation (GGA)\cite{Perdew:96}. Relativistic effects and spin-orbit coupling were included. For \la\/, we found that an on-site Coulomb repulsion of U = 2.08 eV\cite{Anisimov:93} would give good agreement between the Lu 4\textit{f} orbitals to the binding energies of the core levels measured in experiment. Apart from the position of the filled $4f$ core levels, the $U$ = 0 and $U$ = 2.08 DFT calculations for \la\/ were otherwise nearly identical, and the value of $U$ had no impact on the near-$E_{F}$ electronic structure. For \ya\/, calculations were performed both with and without application of $U$ to the Yb 4\textit{f} orbitals, which was found to have a significant impact on the near-E$_{F}$ electronic structure. (Fig. S4)

\subsection{RXES}

RXES spectra were collected at the Cornell High Energy Synchrotron Source (CHESS) at the C1 bend magnet beamline under ring conditions of 5.3 GeV and 100 mA. Incident x-ray radiation was monochromated using a Rh mirror and a sagittal focus double Si(2 2 0) crystal monochromator. The incident energy was calibrated using a Cu foil. The x-ray emission was monochromated and focused using five spherically bent Ge(6 2 0) crystals in the Rowland geometry by using the CHESS dual array valence emission spectrometer (DAVES) \cite{Finkelstein:16}. X-rays were finally collected with a Pilatus 100K area detector (Dectris). Use of an area detector offered significant advantages for the current experiment in terms of ease of alignment and reliable background subtraction. Two regions of interest (ROIs) were chosen, one containing more than 95$\%$ of the emission signal and another centered on the first ROI, but four times in size. The larger ROI was used to correct for the average background counts as

\begin{equation}
    I_{corrected} = I_{ROI_{1}} - Area_{ROI_{1}}\times(\frac{I_{ROI_{2}} - I_{ROI_{1}}}{Area_{ROI_{2}} - Area_{ROI_{1}}})
\end{equation}

Measured counts were further corrected for variations in incident photon flux by normalizing with the measured incident flux using a N$_{2}$-filled ionization chamber placed upstream of the sample stage. X-ray emission energy was calibrated measuring the K$_{\alpha1}$ and K$_{\alpha2}$  lines of a Cu foil. The overall energy resolution of the setup was determined to be better than $\approx$ 3 eV measuring quasi-elastic scattering from a polyimide sample. To minimize photodamage, a fast shutter was placed upstream of the ionization chambers that would only open during active data taking, thus minimizing x-ray dosage of the samples. No photodamage was observed even after taking more than 4 scans (the maximum number of scans used for measurements at a particular spot) at a single spot. The sample was mounted on a closed cycle cryostat with base temperature of 45 K. A helium filled bag covering most of the x-ray path between the sample, analyzer, and detector was placed to reduce air attenuation along the x-ray path.

 
\makeatletter 
\def\tagform@#1{\maketag@@@{(S\ignorespaces#1\unskip\@@italiccorr)}}
\makeatother
\setcounter{equation}{0}

\section{Supplementary Information}

\subsection{Determination of k$_{z}$}

Our ARPES measurements access a two-dimensional projection of the three-dimensional Brillouin zone centered around a k$_{z}$ value that depends on the incident photon energy.
\begin{equation}
    k_{z} = \sqrt{\frac{2m}{\hbar^{2}}(E_{h\nu} + V_{0} - \phi) - k_{||}^{2}}
\end{equation}
where, E$_{h\nu}$ is the incident photon energy, (V$_{0}$ - $\phi$) is the inner potential, and k$_{||}$ is the in-plane momentum.
Both \ya\/ and \la\/ have a three-dimensional cubic structure and therefore, are expected to have strong dispersion in their electronic band structure along k$_{z}$. We estimate k$_{z}$ by comparing the measured band structure of \la\/ with the results from density functional theory (DFT) calculations. Such an approach, however, poses a challenge for \ya\/ as it is well known that DFT+U methods perform poorly for Kondo lattice systems because they fail to satisfactorily account for the strong correlations of the partially filled \textit{f} shell \cite{Zwicknagl:93}. Nevertheless, such a method should perform reasonably well for \la\/, which is a conventional metal system with a filled \textit{f} shell.
In Fig. 1a-c, of the main text and in the  Fig. S1 we show that the simulated two-dimensional Fermi surface topology at k$_{z}$ = $\Gamma$ and the corresponding band dispersion along (0, 0 - 0, $\pi$) are in good agreement with the measured Fermi surface and ARPES spectra, respectively, obtained using a photon energy of 21.2 eV. Furthermore, comparing our data against calculated Fermi surface contour plots and band dispersions for different k$_{z}$ values we establish that the best agreement is obtained for k$_{z}$ $\approx$ $\Gamma$ (Fig. S1b-d). Similar exercise for data collected using 40.8 eV photons leads us to conclude that we are probing a two-dimensional momentum region centered at k$_{z}$ $\approx$ X (Fig. S1b,e-f). 
A Fermi surface map taken with a photon energy of 40.8 eV in \la\/ is shown in Fig. S1e (right panel), where the electron pocket centered at ($\pi$, $\pi$) is closely reproduced by the calculation, with exception of the star-shaped hole like feature at X, which we ascribe to weak photoemission matrix elements. Nevertheless, comparing measured band dispersions along a momentum cut shown in Fig. S1b (blue line) with our DFT calculations we also obtain the best agreement for k$_{z}$ = X.

Similar band dispersions are observed in \ya\/ for broadly dispersive, primarily Al-derived bands at higher binding energies for identical momentum cuts and photon energies, as shown in Fig. 1f (main text), indicating we are probing identical k$_{z}$ region with a particular photon energy in both \la\/ and \ya\/. This correspondence can be seen in many different \textit{E} vs. \textit{k} plots taken at different momentum regions of the Brillouin zone, shown in Fig. S3d-k, assuming a shift in chemical potential due to different average rare-earth valence in these two systems. In fact, a similar two-dimensional electronic structure over the whole Brillouin zone is observed for \la\/ at 0.95 eV binding energy and \ya\/ at 0.22 eV binding energy (Fig. S3a,b).

As expected, DFT fails to reproduce the experimental low-energy electronic structure of \ya\/ (within ~ 1 eV), due to the hybridization of the light bands with the renormalized Yb 4\textit{f} as shown in Fig. S2. Nevertheless, the band dispersions at higher binding energies away from the 4\textit{f} states also give the closest agreement with k$_{z}$ = $\Gamma$ and k$_{z}$ = X with measurements using He I$\alpha$ and He II$\alpha$ photons, respectively, as also found in its uncorrelated analogue \la\/. Therefore, the experimentally observed correspondence in the electronic structures of \la\/ and \ya\/, along with a detailed comparison with our DFT calculations establishes that we are probing identical k$_{z}$ values for both \la\/ and \ya\/, which is k$_{z}$ = $\Gamma$ for He I$\alpha$ (21.2 eV) and k$_{z}$ = X for He II$\alpha$ (40.8 eV) photon energies, respectively. Using equation 1, from the estimated k$_{z}$ values corresponding to the excitation energies of 21.2 eV and 40.8 eV, we obtain the inner potential of the (001) \la\// \ya\/ surface, V$_{0}$ - $\phi$ = 13.66 eV.

\subsection{Determination of Yb valence in YbAl$_{3}$}

Temperature-dependent Yb valence in \ya\/ is determined by x-ray photoemission spectroscopy (XPS) and resonant x-ray emission spectroscopy (RXES).
In the case of a mixed-valence system, such as \ya\/, x-ray photoemission spectrum is expected to consist of two different final state multiplets corresponding to different initial valence configurations. For \ya\/, both 4\fo\/ (spectral peaks between 0 - 2 eV) and 4\ft\/ (spectral peaks between 5 - 12 eV) final states are observed, separated by U$_{ff}$ $\approx$ 6.5 eV, shown in Fig. S3a that correspond to Yb$^{2+}$ 4$f^{14}$ and Yb$^{3+}$ 4$f^{13}$ initial states, respectively. The average Yb valence $\nu$ is determined by evaluating the integral spectral weight corresponding to 4$f^{13}$ (I$^{2+}$) and  4$f^{12}$ (I$^{3+}$) final states as\cite{Moreschini:07,Kummer:11}

\begin{equation}
\nu = 3 - \frac{I^{2+}/14}{I^{2+}/14 + I^{3+}/13}
\end{equation}

In Fig. S5b, we show the estimated variation in the average Yb valence in \ya\/ as a function of sample temperature. It should be noted that the measured photoemission intensity can be strongly influenced by matrix element and final state effects. Furthermore, to have a good estimation of the background intensity, knowledge of the non-4\textit{f} spectral weight at that particular photon energy and contributions from inelastically scattered electrons is required, making the estimation process tricky. Indeed, there is a large distribution in the reported absolute Yb valence in \ya\/, dependent primarily on the nature of the probe. Nonetheless, a relative change in intensity as a function of temperature can be determined much more precisely leading to an accurate determination of temperature-dependent changes in the Yb valence, because a change in temperature is expected to negligibly affect the factors modulating measured photoemission intensity and non-4\textit{f} spectral weight, which is shown in Fig. 2c of the main text.\\

RXES is a bulk-sensitive photon-in photon-out technique that can be used as an alternative to XPS for determination of rare-earth valence in Kondo lattice systems. The incident photon energy is tuned to the Yb L$_{III}$ absorption edge, thus resonantly enhancing the signal from Yb in \ya\/, making it element specific. Ground states with Yb$^{2+}$ and Yb$^{3+}$ valence states undergo slightly different transitions resulting in the corresponding absorption edges being separated in energy by $\approx$ 8 eV in \ya\/ \cite{Moreschini:07,Kummer:11,Kumar:08}. Measurement of the absorption spectral weight corresponding to different Yb valence configurations leads to an estimate of the average Yb valence. In RXES, the emitted x-rays are energy resolved and only a particular decay channel is collected, resulting in greater resolving power compared to x-ray absorption spectroscopy (XAS) \cite{Moreschini:07,Kummer:11,Glatzel:05}. 

In Fig. S6c we show normalized spectra taken at the L$_{\alpha1}$ emission line both at 300 K and 45 K. It is clearly seen that the contribution of the Yb$^{2+}$ absorption peak increases at the lower temperature indicating that the average Yb valence decreases with temperature. To obtain a quantitative estimate of the average Yb valence, RXES spectra were fitted with Voigt functions for absorption peaks corresponding to Yb$^{2+}$ and Yb$^{3+}$. Absorption features corresponding to Yb$^{3+}$ exhibit a double-peak structure requiring a two-component Voigt function. The double-peak structure for the Yb$^{3+}$ component has been seen in other Yb intermetallic compounds such as YbCuAl\cite{Yamaoka:13} and Yb$_{2}$Ni$_{12}$P$_{7}$\cite{Jiang:15}, and is ascribed to the crystal field splitting of the unoccupied Yb 5\textit{d} states. A series of arctan-like functions capturing the edge jumps corresponding to the absorption edges is used to estimate the background. The average Yb valence is estimated using the following formula

\begin{equation}
\nu = 2 + \frac{I(3^{+})}{I(2^{+}) + I(3^{+})}
\end{equation} 
                                
where, I(2$^{+}$) and I(3$^{+}$) are respective intensities of Yb$^{2+}$ and Yb$^{3+}$ components \cite{Kummer:11,Yamaoka:13}, and is found to be 2.83 $\pm$ 0.01 at 300 K and 2.78 $\pm$ 0.01 at 45 K. Thus, from 300 K to 45 K the average Yb valence in \ya\/ decreases by $\approx$ 0.05, consistent with results from similar measurements on \ya\/ single crystals\cite{Moreschini:07,Suga:05,Bauer:04,Lawrence:94}.

\subsection{Fits to YbAl$_{3}$ data}

Energy distribution curves (EDCs) at different temperatures, as shown in Fig. S7a, are fitted with a functional form consisting of Lorentzians multiplied by the corresponding Fermi-Dirac distribution and convoluted with a Gaussian representing instrumental broadening, the full width at half maximum (FWHM) of which is estimated measuring a gold reference. A Shirley background that takes into account contribution from the inelastically scattered electrons is subtracted before the fitting process \cite{Shirley:72}.
Similar fits have also been performed to extract crystalline electric field (CEF) split states for the EDCs shown in Fig. S8b,c.  
Momentum distribution curves (MDCs) at the Fermi energy, shown in Fig. S7b, are fitted with a functional form consisting of two Lorentzians and a linear background. The MDC at E$_{F}$ taken at 21 K, in contrast to those taken at higher temperatures, exhibits only a single peak and hence, is fitted with a single Lorentzian. This further establishes that at 21 K the electron-like pocket centered at (0, 0) is completely lifted above the Fermi level and only the residual spectral weight is observed. 
To access thermally occupied states above the Fermi level, the \textit{E} vs. \textit{k} plot obtained is divided by the corresponding resolution broadened Fermi-Dirac distribution, shown in Fig. S7c, where the band bottom of the electron pocket is seen above the Fermi level.

\begin{acknowledgments}
We thank Yang Liu, J.W. Allen, J.D. Denlinger, G.A. Sawatzky and H. Takagi for helpful discussions. This work was supported by the National Science Foundation through DMR-0847385 and the Materials Research Science and Engineering Centers program (DMR-1120296, the Cornell Center for Materials Research), the Research Corporation for Science Advancement (2002S), and by the Gordon and Betty Moore Foundation as part of the EPiQS initiative (GBMF3850). Support from the Air Force Office of Scientific Research was through FA2386-12-1-3013. This work was performed in part at the Cornell NanoScale Facility, a member of the National Nanotechnology Infrastructure Network, which was supported by the National Science Foundation (Grant No. ECCS- 0335765). This work was also performed in part at the Cornell High Energy Synchrotron Source (CHESS) which is supported by the National Science Foundation and the National Institutes of Health/National Institute of General Medical Sciences under NSF award DMR-1332208. H.I.W. and J.P.R. acknowledge support from the NSF Integrative Graduate Education and Research Traineeship program (DGE-0903653), and H.I.W. also acknowledges support from the NSF Graduate Research Fellowship (DGE-1144153). 
\end{acknowledgments}

\section*{Author contribution}
S.C. and K.M.S. conceived the idea. Thin film growth, film characterization, ARPES, XPS and DFT calculations were performed and analyzed by S.C. RXES was performed by S.C., J.P.R., and H.I.W. with assistance and input from K.D.F, and analyzed by S.C. The manuscript was prepared by S.C. and K.M.S. D.G.S. and K.M.S. supervised the study. All authors discussed results and commented on the manuscript.

\section*{Additional information}
The authors declare no competing financial interests. Supplementary information accompanies this paper. Correspondence and requests for materials should be addressed to K.M.S.\\

\newpage


\begin{figure}
\includegraphics[width=1\columnwidth]{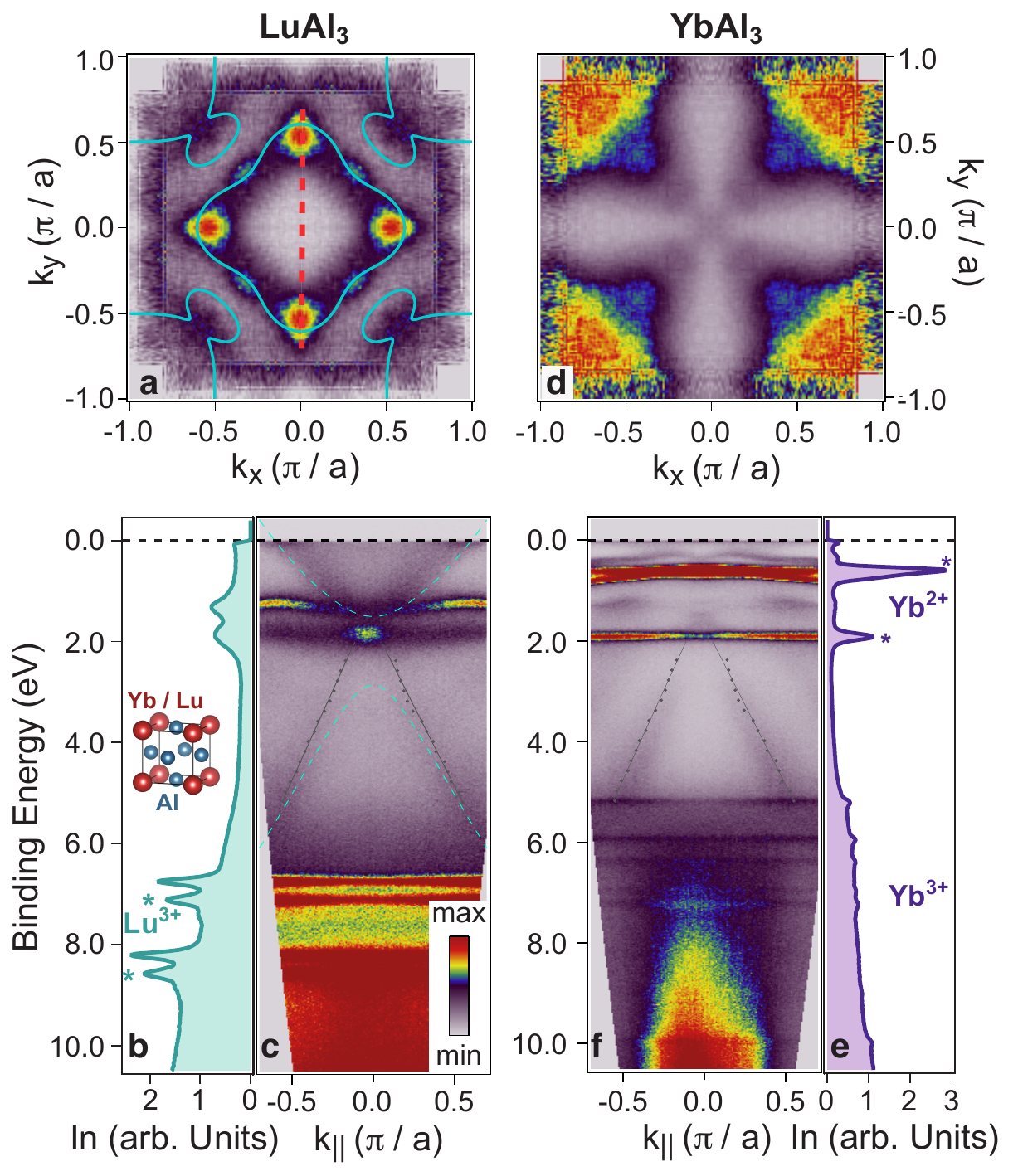}
\caption{\textbf{Electronic Structure and Fermi Surfaces of YbAl$_{3}$ and LuAl$_{3}$.} Fermi surface maps and Energy Distribution Curves (EDCs) and $E$ vs. $k$ dispersion for \textbf{a - c,} \la\/ and \textbf{d - f,} \ya\/, all measured with $h\nu$ = 21.2 eV at 21 K. Experimental Fermi surfaces of \textbf{a,} \la\/ and \textbf{d,} \ya. DFT calculations (green lines) of the Fermi surface topology at $k_z$ = 0 for \la\/ with $U$ = 0 are overlaid in \textbf{a.} Momentum-integrated EDCs of  \textbf{b,} \la\/ and  \textbf{e,} \ya, with surface core levels marked as asterisks. $E$ vs. $k$ dispersions for \textbf{c,} \la\/ and \textbf{f,} \ya, together with DFT calculations of the band structure in \la\/ (green) shown in \textbf{c.} The similarity between the dispersion of the light band between 2-6 eV in \ya\/ and \la\/ suggest that both compounds have similar inner potentials, and that both measurements are at $k_{z} = 0 \pm 0.1 \pi/a$.}
\label{fig:ES}
\end{figure}

\begin{figure*}[t]
\includegraphics[width=1\textwidth]{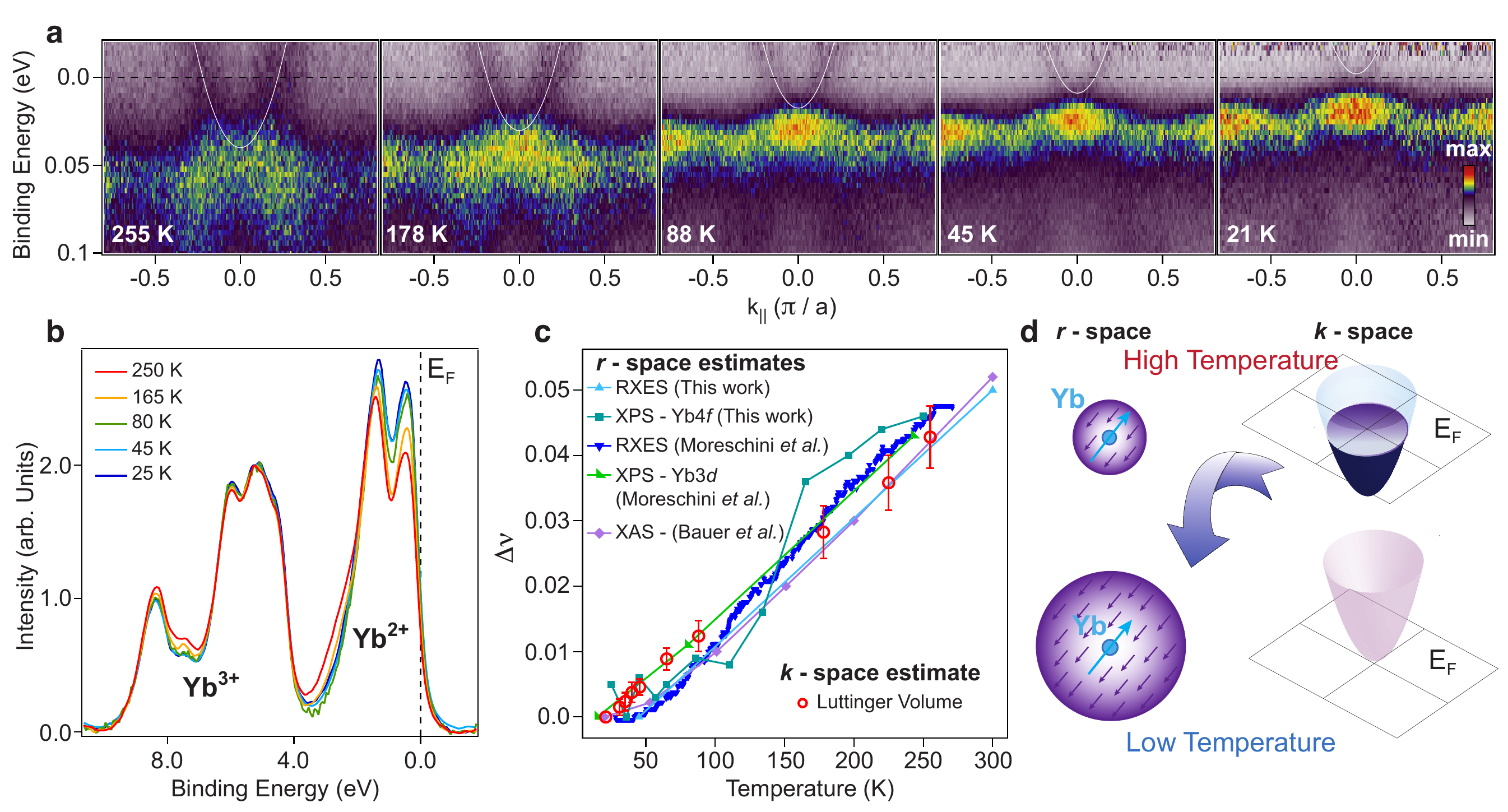}
\caption{\textbf{Correspondence between \textit{r}-space and \textit{k}-space electronic structure in YbAl$_{3}$.} \textbf{a,} Evolution of the low energy electronic structure with temperature. $E$ vs. $k$ dispersions are divided by the corresponding resolution-broadened Fermi-Dirac distribution to emphasize thermally occupied states above $E_{F}$. White lines are guides to the eye showing evolution of the electron-like pocket centered at (0, 0, 0). \textbf{b,} XPS spectra showing the temperature-dependent intensity variation of the 4\fo\/ and 4\ft\/ final states in \ya, with Shirley background subtraction \cite{Shirley:72} and normalized by the 4\ft\/ final state intensity. \textbf{c,} Temperature dependence of the change in Luttinger volume, estimated from the size of the electron pocket at (0, 0, 0) and of the change in Yb valence, measured by core level spectroscopy, revealing a precise one-to-one correspondence. \textbf{d,} Schematic illustrating the temperature-dependent relationship between \textbf{$r$}-space and \textbf{$k$}-space electronic structure in YbAl$_{3}$.}
\label{fig:krcorrespondence}
\end{figure*}

\begin{figure}
\includegraphics[width=1\columnwidth]{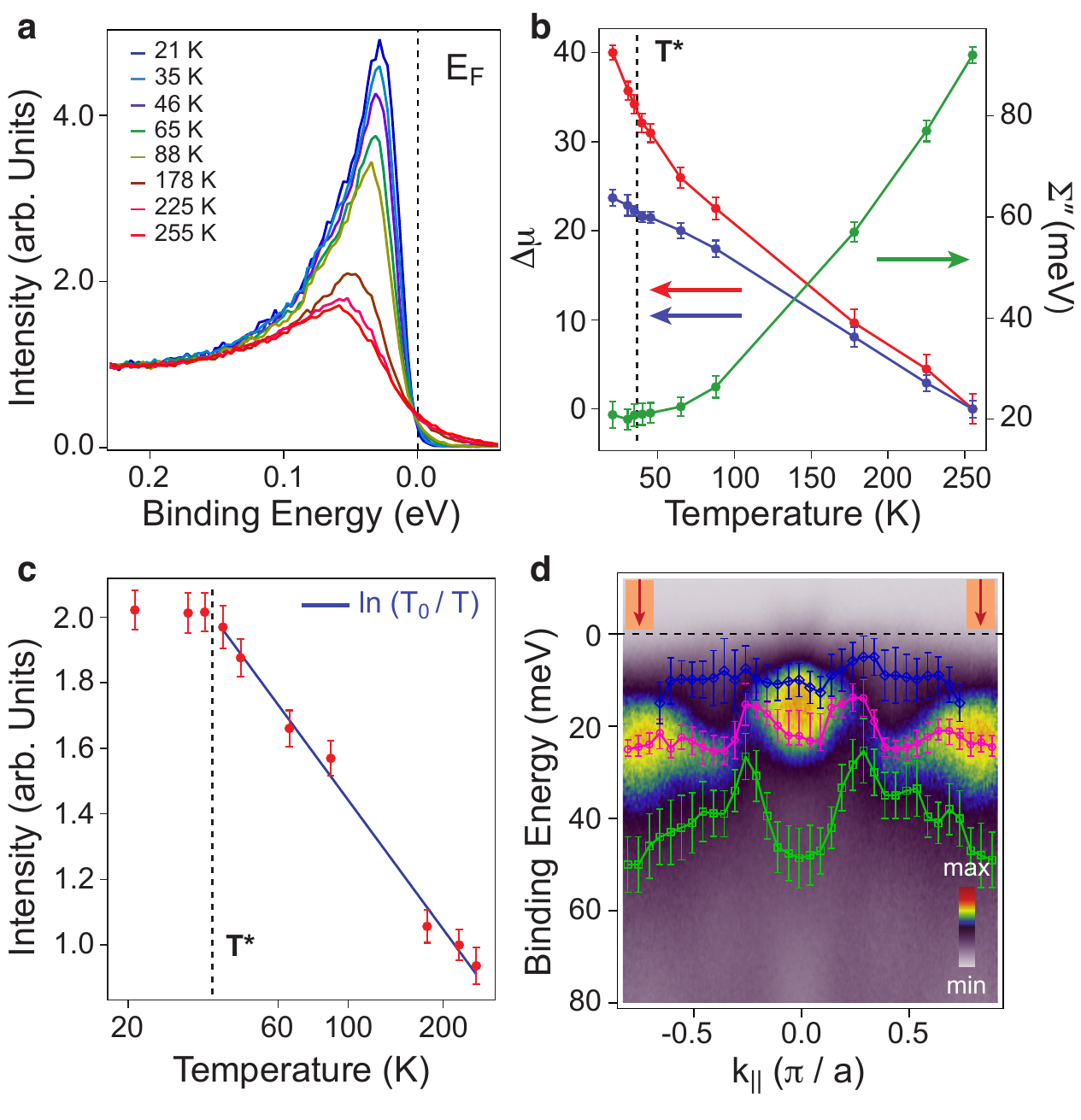}
\caption{\textbf{Evolution of the Yb 4\textit{f} states and crystal field effects in YbAl$_{3}$}\/ \textbf{a,} Evolution of the Kondo resonance peak with temperature, from integrating EDCs over $k$ region highlighted as red in panel \textbf{d}. \textbf{b,} Change in the chemical potential ($\Delta \mu$) and in the 4\textit{f} quasiparticle scattering rate with temperature. $\Delta \mu$ is estimated from the shift in binding energy of the 4\textit{f} derived heavy band (blue) and the band bottom of the light electron-like pocket at (0, 0) (red) relative to 255 K. \textbf{c,} Temperature dependence of the integrated spectral weight (0-0.2 eV) of the 4\textit{f} states which show a $ln\frac{T_{0}}{T}$ behavior above T$^{*}$ = 37 K. \textbf{d,} High resolution $E$ vs. $k$ plot along (0, 0) - (0, $\pi$) at 21 K showing dispersive crystal electric field (CEF) split states. Extracted dispersions of the three different CEF split states (shown in blue, pink and green) are superimposed on the image plot.}
\label{fig:Coherence}
\end{figure}


\renewcommand{\thefigure}{S\arabic{figure}}
\setcounter{figure}{0} 
\makeatletter 
\def\tagform@#1{\maketag@@@{(S\ignorespaces#1\unskip\@@italiccorr)}}
\makeatother
\setcounter{section}{0}
\setcounter{equation}{0}


\begin{figure*}
\includegraphics[width=1\textwidth]{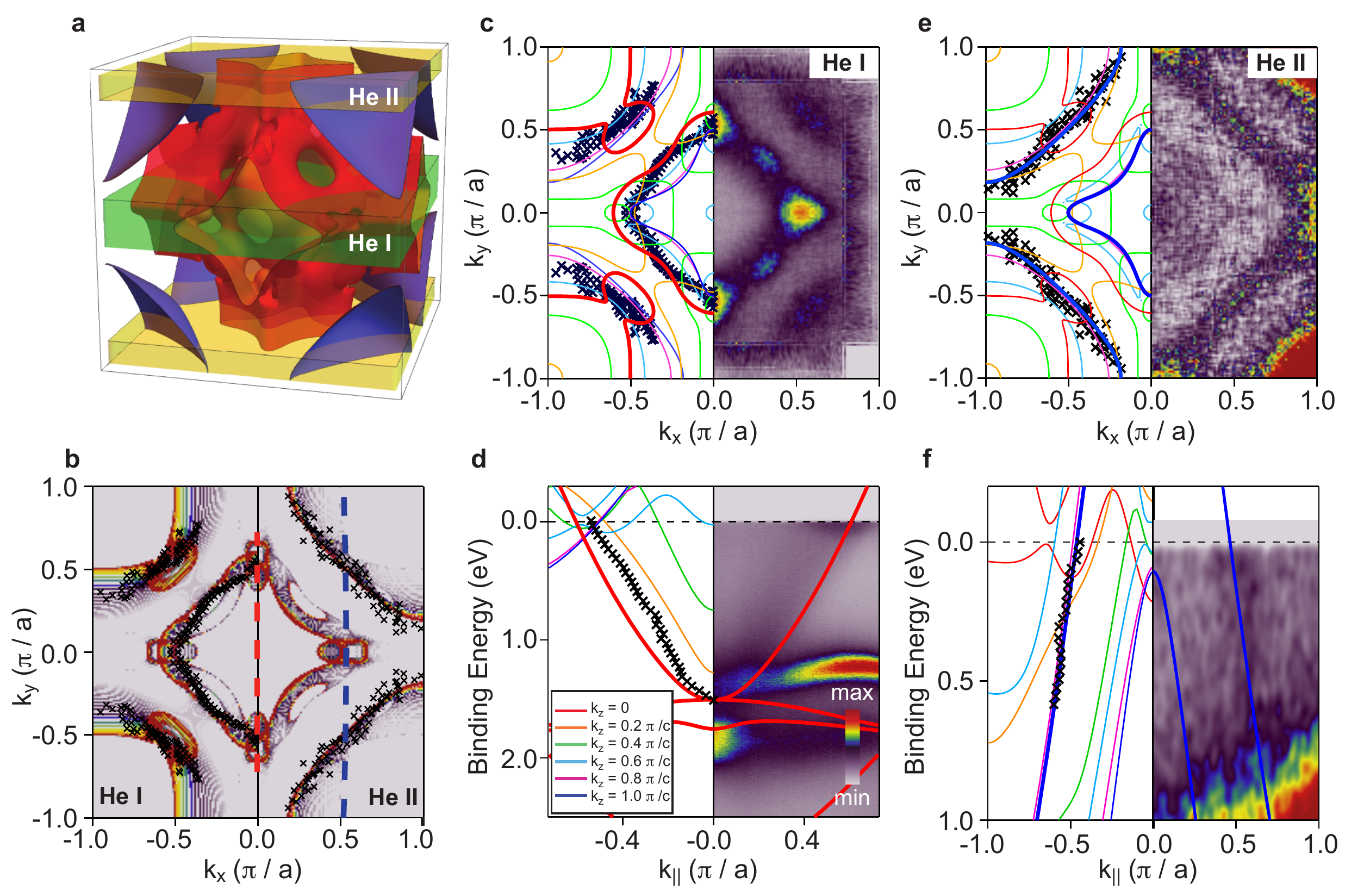}
\caption{\textbf{Determination of k$_{z}$ in LuAl$_{3}$} \textbf{a,} DFT-calculated three-dimensional Fermi surface of \la\/ with U = 0. Momentum regions centered at k$_{z}$ = $\Gamma$ accessed by He I$\alpha$ (21.2 eV) and at k$_{z}$ = X accessed by He II$\alpha$ (40.8 eV) photon energies are shown by green and yellow regions, respectively. \textbf{b,}Simulated two-dimensional Fermi surface maps from DFT calculations with k$_{z}$ smearing taken over the thickness of the slabs shown in \textbf{a,} and overlaid with extracted k$_{F}$s from experiment are shown for both He I$\alpha$ and He II$\alpha$. Measured Fermi surface map and corresponding k$_{F}$s along with calculated Fermi surface contour plots for different k$_{z}$ values for \textbf{c,} 21.2 eV and \textbf{e,} 40.8 eV photon energies. Fermi surface contour plots corresponding to k$_{z}$ = 0 and k$_{z}$ = $\pi$/c are shown in bold in \textbf{c,} and \textbf{e,} emphasizing good agreement with the extracted k$_{F}$s using He I$\alpha$ and He II$\alpha$ photons, respectively. \textit{E} vs. \textit{k} dispersion, extracted dispersion and calculated dispersions for different k$_{z}$ values \textbf{d,} for He I$\alpha$, corresponding to the momentum line cuts shown in red in \textbf{b}. \textbf{f,} for He II$\alpha$, corresponding to the momentum line cuts shown in blue in \textbf{b}. Extracted dispersion from the \textit{E} vs. \textit{k} plots in \textbf{d,} and \textbf{f,} are in good agreement with the calculated dispersions that are shown in bold for k$_{z}$ = 0 and k$_{z}$ = $\pi$/c, respectively. For clarity, dispersions for only those bands that form the Fermi surface are shown for k$_{z}$ values different from k$_{z}$ = 0 and k$_{z}$ = $\pi$/c in the left panels of \textbf{d,} and \textbf{f,} respectively; some of the bands that remain completely below E$_{F}$, which can be seen in ARPES experiment, are omitted from the plotted bands on the left.}
\label{fig:kz_LA}
\end{figure*}

\begin{figure*}
\includegraphics[width=1\textwidth]{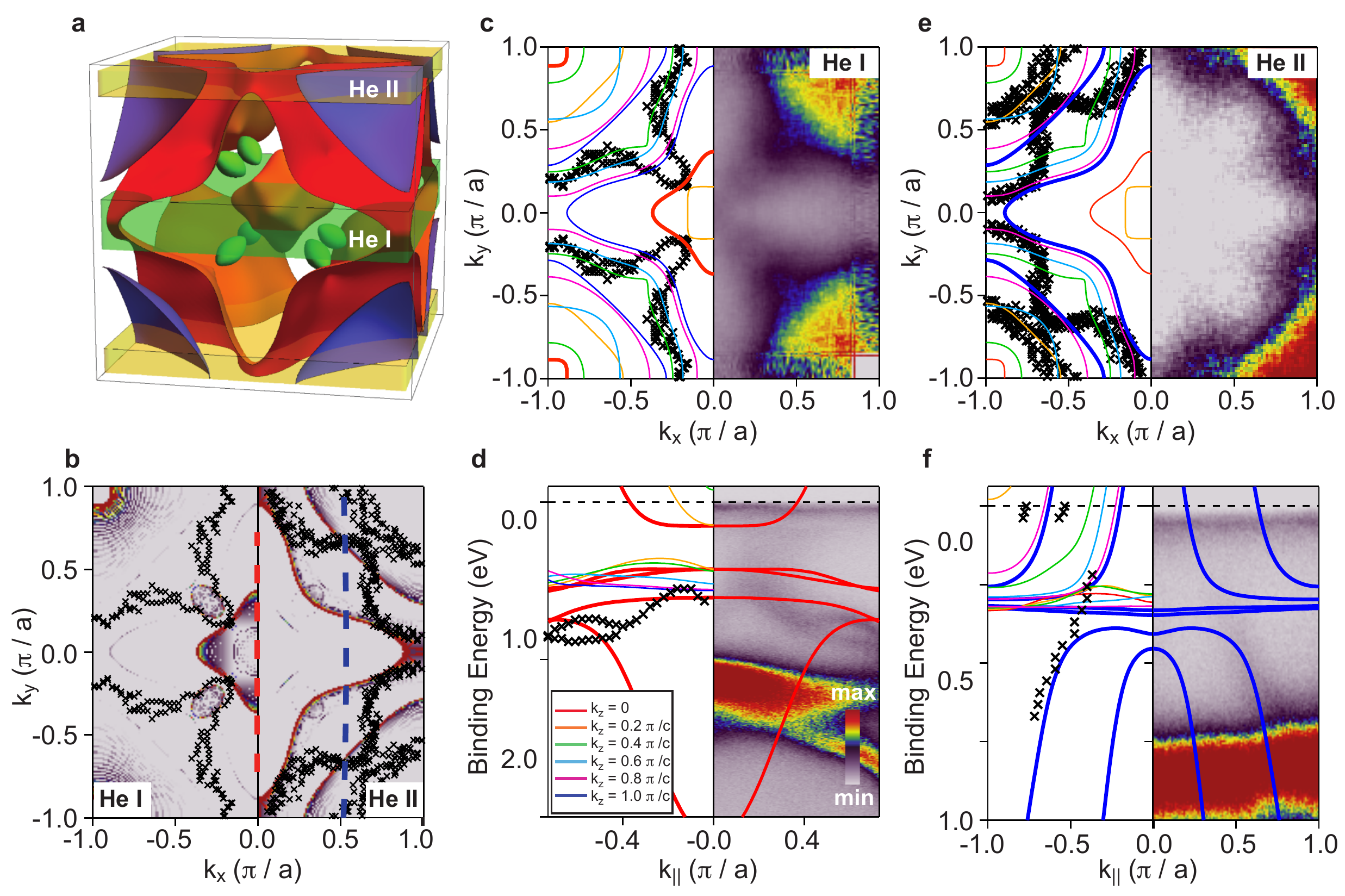}
\caption{\textbf{Calculated and measured electronic structure in YbAl$_{3}$} \textbf{a,} DFT-calculated three-dimensional Fermi surface of \ya\/ with U = 0. Momentum regions centered at k$_{z}$ = $\Gamma$ accessed by He I$\alpha$ (21.2 eV) and at k$_{z}$ = X accessed by He II$\alpha$ (40.8 eV) photon energies are shown by green and yellow regions, respectively. \textbf{b,} Projected two-dimensional Fermi surface maps centered at k$_{z}$ = $\Gamma$ (He I$\alpha$) and k$_{z}$ = X (He II$\alpha$) over the momentum regions shown in \textbf{a}. Fermi wave vectors (k$_{F}$s) extracted from photoemission measurements using He I$\alpha$ and He II$\alpha$ are overlaid on top. Measured Fermi surface map and corresponding k$_{F}$s along with calculated Fermi surface contour plots for different k$_{z}$ values for \textbf{c,} 21.2 eV and \textbf{e,} 40.8 eV photon energies. Fermi surface contour plots corresponding to k$_{z}$ = 0 and k$_{z}$ = $\pi$/c are shown in bold in \textbf{c,} and \textbf{e,} respectively. These all show generally poor agreement, due to the inadequate treatment of the strong correlations of the Yb 4\textit{f} orbitals in the DFT calculation. \textit{E} vs. \textit{k} plot, extracted dispersion and calculated dispersions for different k$_{z}$ values \textbf{d,} for He I$\alpha$, corresponding to the momentum line cuts shown in red in \textbf{b}. \textbf{f,} for He II$\alpha$, corresponding to the momentum line cuts shown in blue in \textbf{b}. For clarity, dispersions for only those bands that form the Fermi surface are shown for k$_{z}$ values different from k$_{z}$ = 0 and k$_{z}$ = $\pi$/c in the left panels of \textbf{d,} and \textbf{f,} respectively.}
\label{fig:kz_YA}
\end{figure*}

\begin{figure*}[t]
\includegraphics[width=1\textwidth]{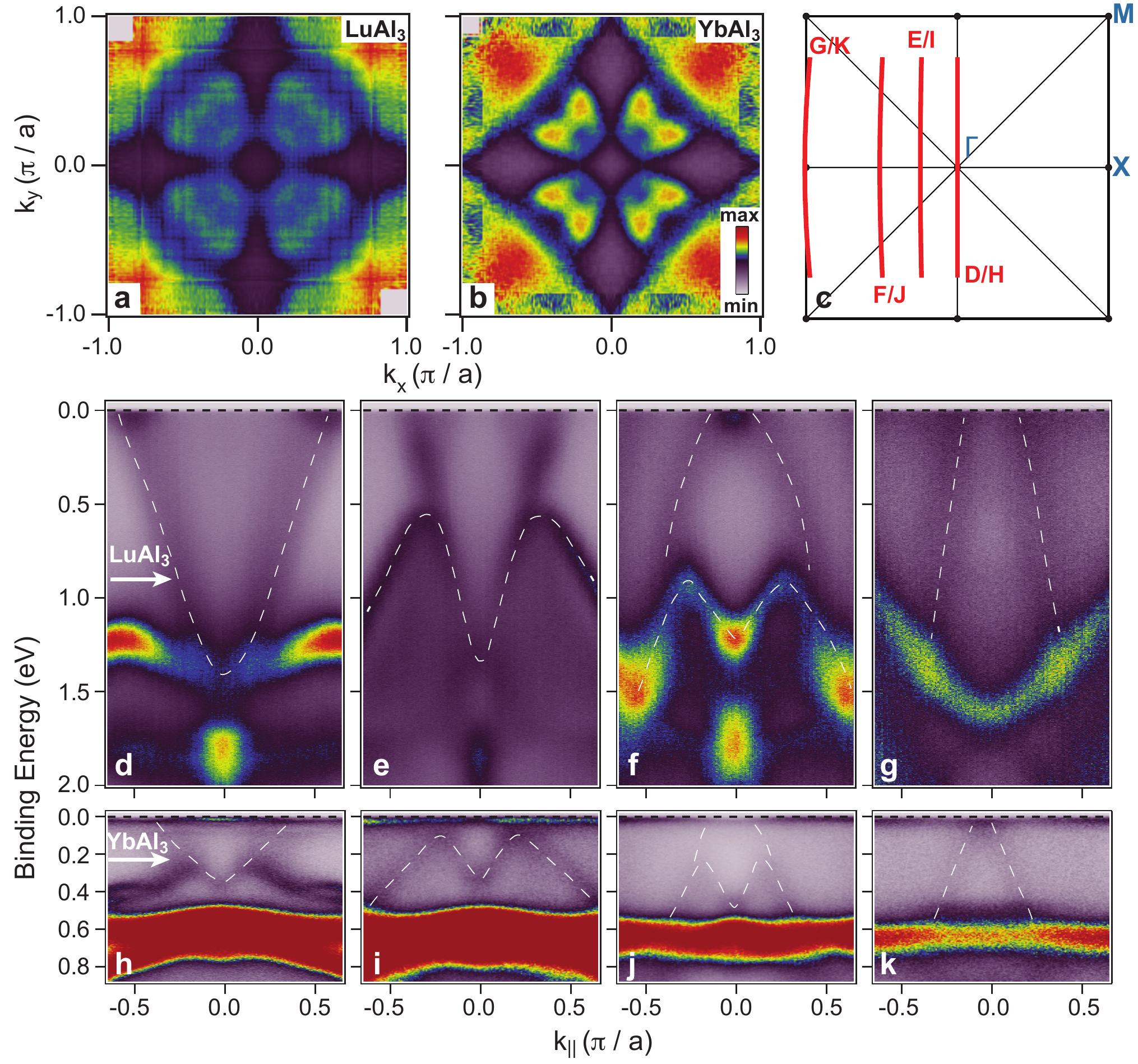}
\caption{\textbf{Similarity in electronic structure between LuAl$_{3}$ and YbAl$_{3}$}Four-fold symmetrized two-dimensional electronic structure of \textbf{a,} \la\/ at 0.95 eV binding energy and of \textbf{b,} \ya\/ at 0.22 eV binding energy, both obtained with a photon energy 21.2 eV (He I$\alpha$) \textbf{c,} The surface Brillouin zone of \la\/ and \ya\/ showing high symmetry points and location of the ARPES cuts in momentum space. \textit{E} vs. \textit{k} plots corresponding to the ARPES cuts in panel c are shown in \textbf{d-g}, for \la\/ and in \textbf{h-k}, for \ya\/. White dotted lines are guides to the eye highlighting the similarity in band dispersions in \la\/ and \ya\/. White arrows in panels \textbf{d} and \textbf{h} indicate the binding energies at which two-dimensional maps shown in \textbf{a} and \textbf{b} are taken, respectively.}
\label{fig:LAYA}
\end{figure*}

\begin{figure*}[t]
\includegraphics[width=0.7\textwidth]{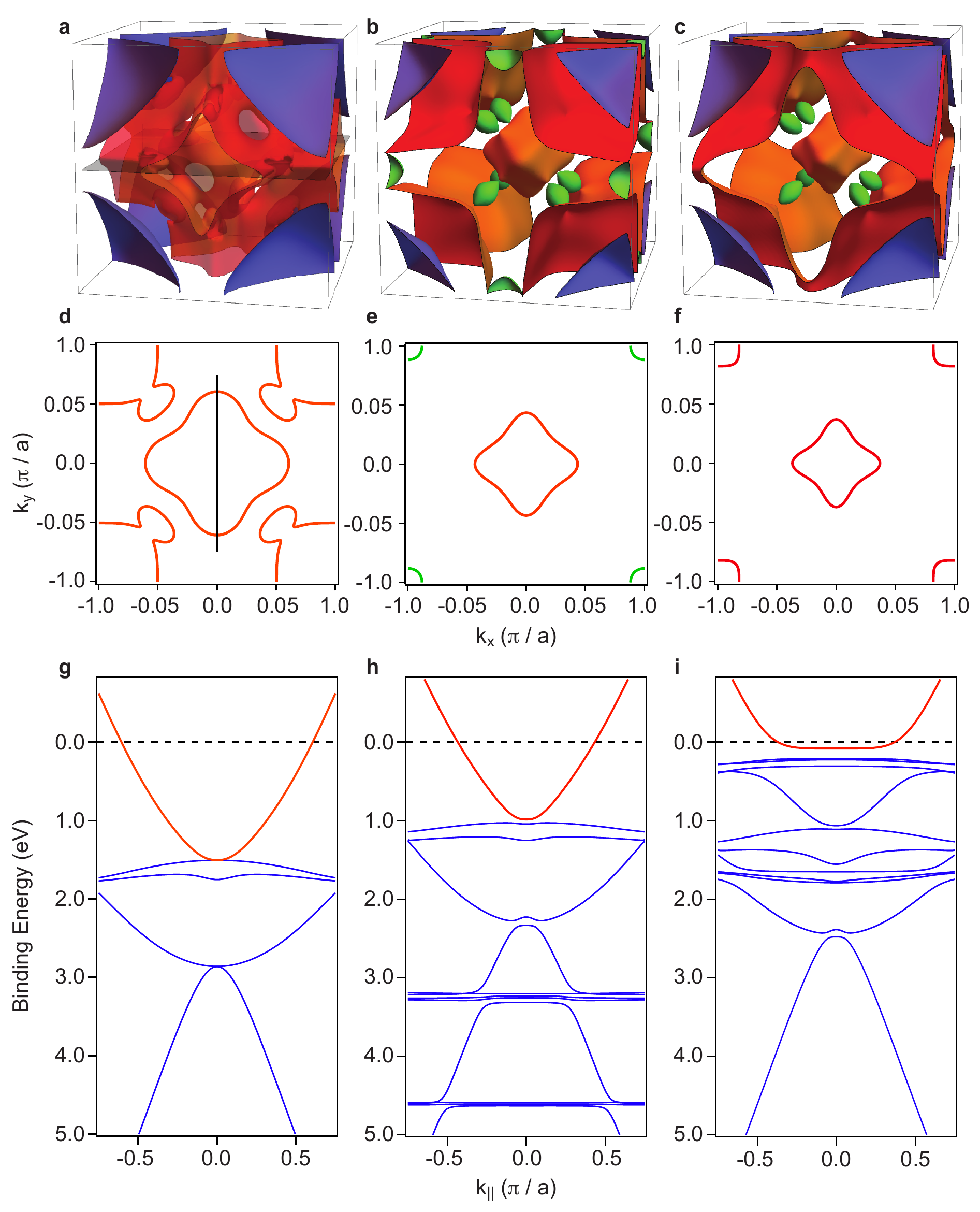}
\caption{\textbf{Evolution of the calculated electronic Fermi surface at $\Gamma$ across LuAl$_{3}$ and YbAl$_{3}$} Calculated three-dimensional Fermi surface for \textbf{a,} \la\/ \textbf{b,} \ya\/ with U = 0 \textbf{c,} \ya\/ with U = 5.46 eV applied to Yb 4\textit{f} orbitals.  Corresponding two-dimensional Fermi surface contour plots at k$_{z}$ = $\Gamma$ for \textbf{d,} \la\/ \textbf{e,} \ya\/ with U = 0 \textbf{f,} \ya\/ with U = 5.46 eV. Corresponding energy-momentum dispersion for a momentum cut (black line) shown in \textbf{d,} for \textbf{g,} \la\/ \textbf{h,} \ya\/ with U = 0 \textbf{i,} \ya\/ with U = 5.46 eV. Dispersion of the electron pocket that forms the quasi-spherical Fermi surface centered at $\Gamma$ is shown in red. It is clearly seen that the binding energy of the Yb 4\textit{f} electronic states strongly influence the size of the quasi-spherical Fermi surface.}
\label{fig:DFT}
\end{figure*}

\begin{figure*}[t]
\includegraphics[width=1\textwidth]{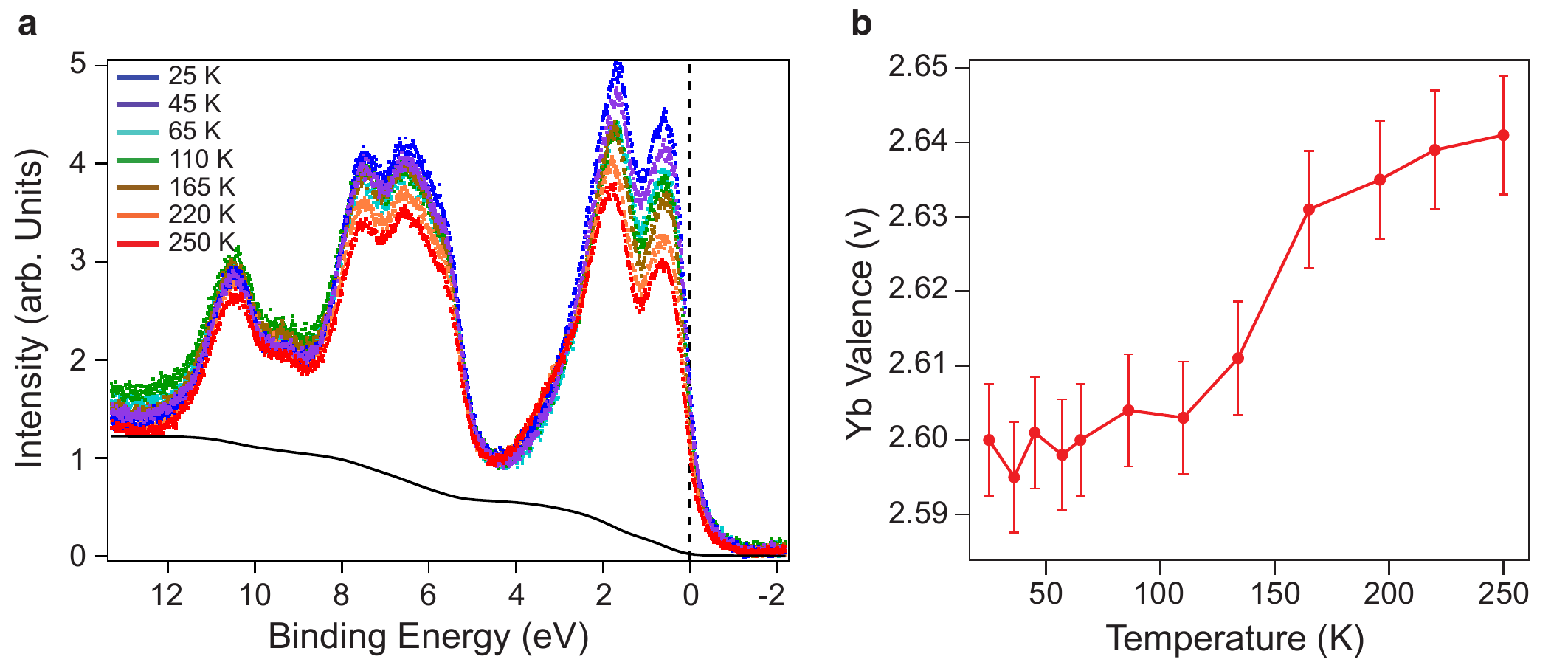}
\caption{\textbf{XPS spectra and Yb valence in YbAl$_{3}$} \textbf{a,} Evolution of XPS spectra with temperature. A Shirley-type background used to estimate contributions from inelastically scattered electrons is shown for the XPS spectra taken at 250 K (black line). \textbf{b,} Yb valence as a function of temperature evaluated estimating spectral intensity corresponding to 4\textit{f}$^{13}$ and 4\textit{f}$^{12}$ final states. Error bars include statistical error of one standard deviation from the fitting process and, also error in the estimated value due to variability in the fit results by holding one and/or more than one peak positions constant during the multi-peak fit.}
\label{fig:XPS}
\end{figure*}

\begin{figure*}[t]
\includegraphics[width=1\textwidth]{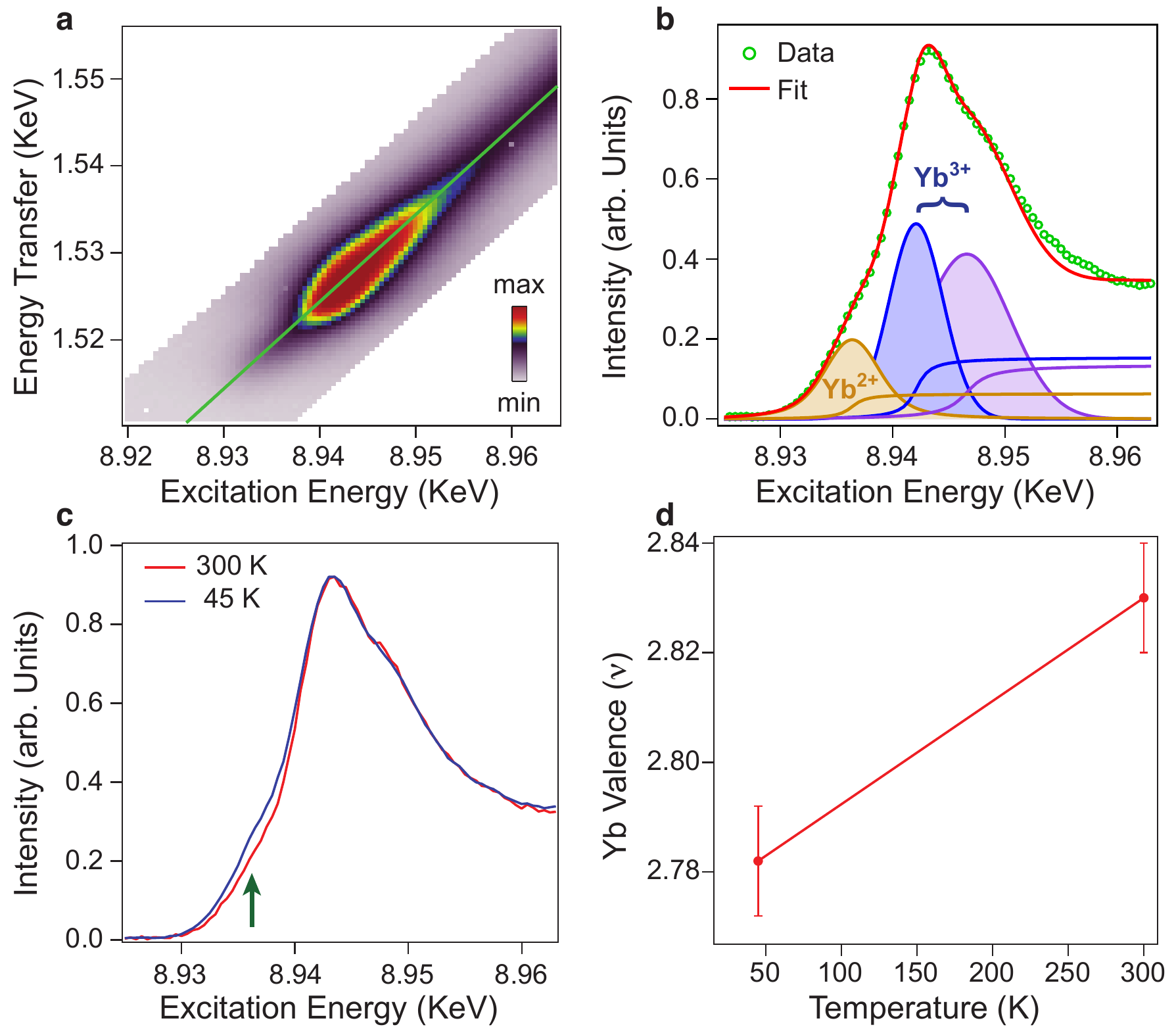}
\caption{\textbf{Resonant x-ray emission spectroscopy (RXES) on YbAl$_{3}$ thin films.}  \textbf{a,} Two-dimensional intensity map of the photon yield around the Yb L$_{\alpha1}$ (7.416 KeV) emission energy plotted as a function of incident energy and energy transfer into the sample (E$_{in}$ - E$_{out}$) The line cut shown in green corresponds to the emission energy L$_{\alpha1}$ \textbf{b,} Intensity variation of the emission spectra with excitation energy at Yb L$_{\alpha1}$  emission energy with the corresponding fit showing contributions from Yb$^{2+}$ and Yb$_{3+}$ components and respective arc-tan like contributions capturing the edge jumps. The absorption feature corresponding to Yb$_{3+}$ has a double peak structure, which is ascribed to the crystal field splitting of the Yb 5\textit{d} band. \textbf{c,} Identical cut as in panel \textbf{b} at two different temperatures. Spectra are normalized to their corresponding maximum in intensity to highlight the enhanced contribution from Yb$^{2+}$ at the lower temperature. \textbf{d,} Estimated Yb valence as a function of temperature as obtained from RXES. Error bars include statistical error of one standard deviation from the fitting process and, also error in the estimated value due to variability in the fit results by holding one / two peak positions constant during the multi-peak fit.}
\label{fig:RXES}
\end{figure*}

\begin{figure*}[t]
\includegraphics[width=1\textwidth]{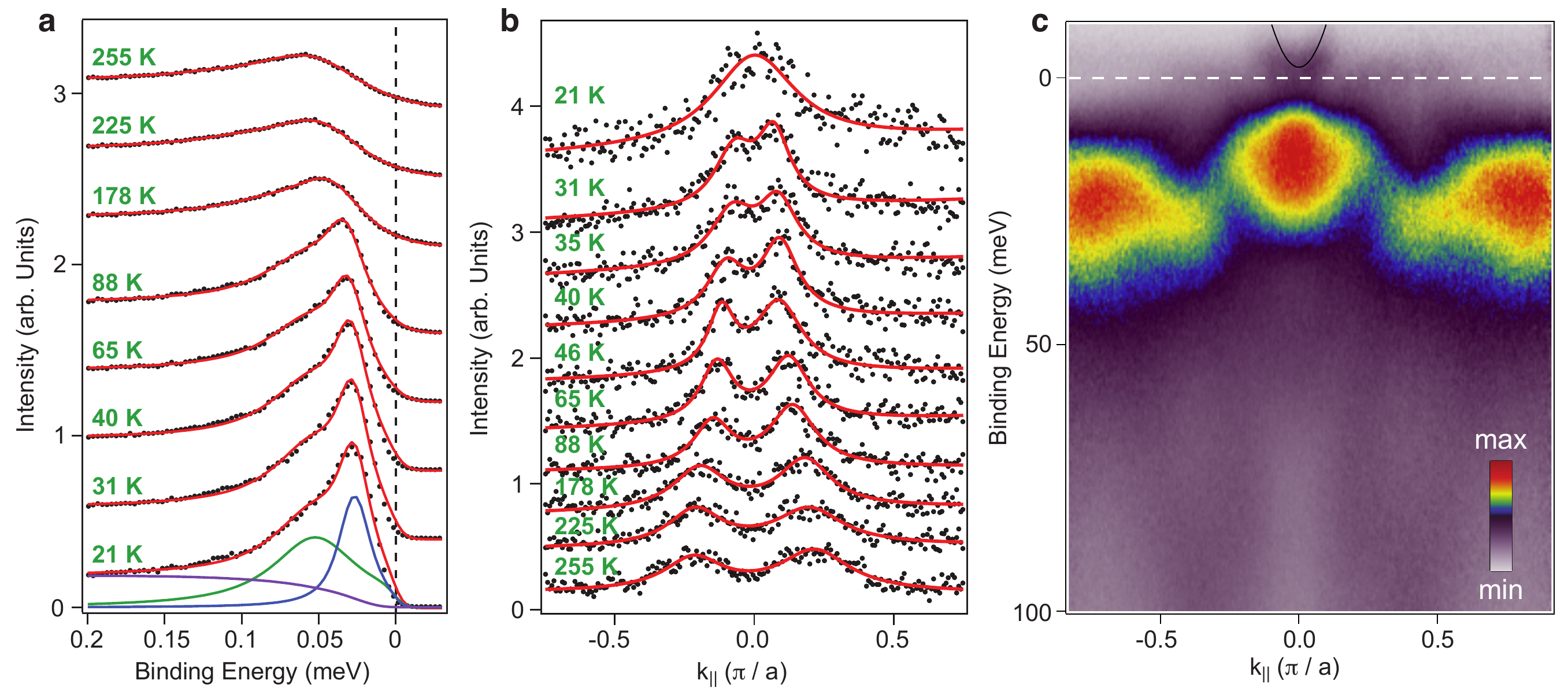}
\caption{\textbf{Fits to YbAl$_{3}$ ARPES data} \textbf{a,} Energy distribution curves (EDCs) at different temperatures obtained after integrating over a momentum region shown as red in Fig. 3d, (main text) along with their corresponding fits (red solid line). Individual contributions to the fitted spectrum (blue and green solid lines) is shown for one such fit. A Shirley background (violet line) has been subtracted out prior to the fitting process that accounts for inelastically scattered electrons. \textbf{b,} Momentum distribution curves (MDCs) at E$_{F}$ taken at different temperatures along with their corresponding fits. \textbf{c,} High resolution, high-statistics \textit{E} vs. \textit{k} plot at 21 K obtained after dividing by the resolution broadened Fermi function. The parabolic band (black solid line), is a guide to the eye for the dispersion of the electron-like pocket at $\Gamma$.}
\label{fig:EDC}
\end{figure*}


\begin{figure*}[t]
\includegraphics[width=1\textwidth]{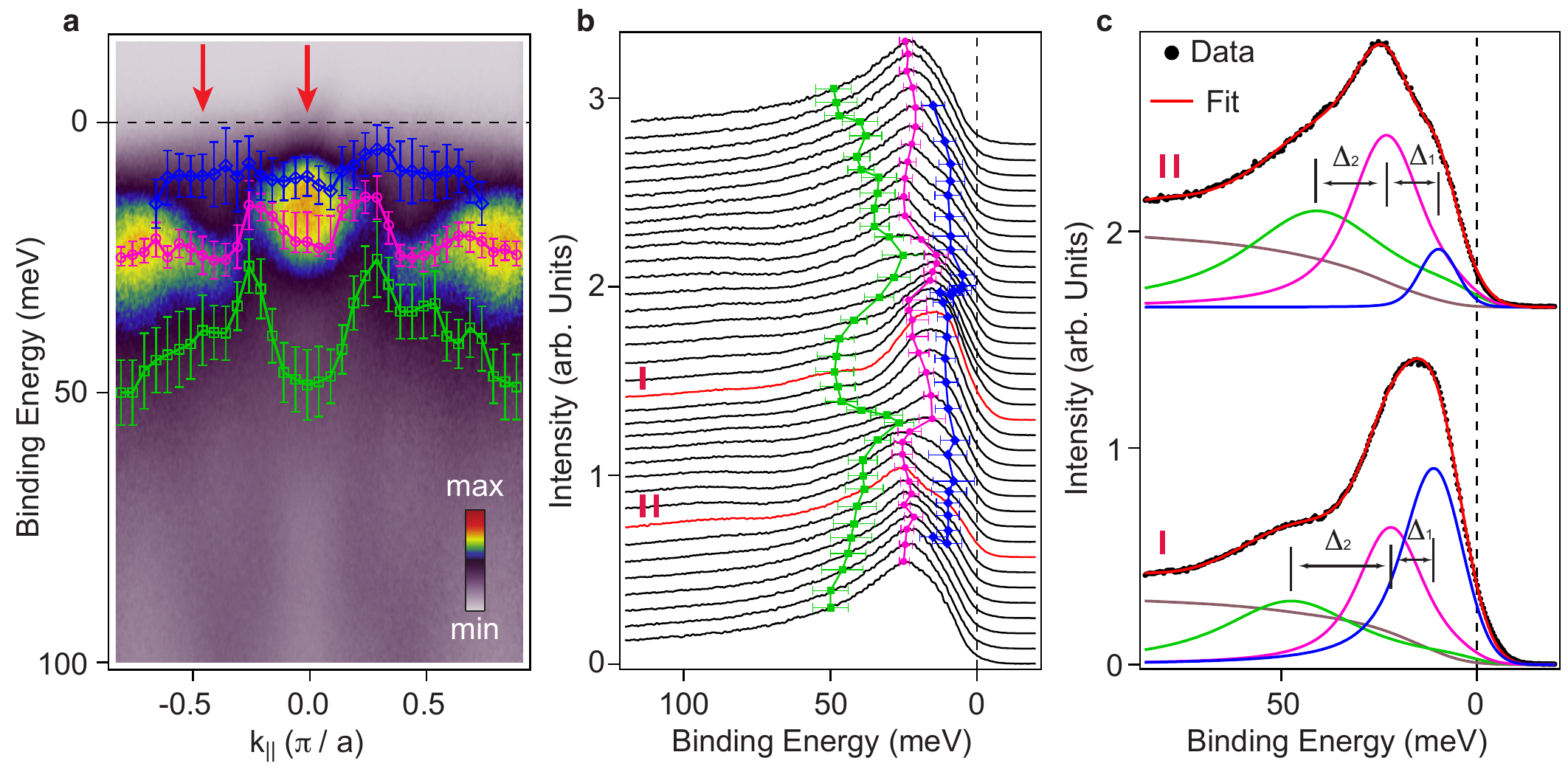}
\caption{\textbf{Crystalline Electric Field (CEF) split states in YbAl$_{3}$} \textbf{a,} High-resolution E vs. k plot along (0, 0) - (0, $\pi$) in \ya\/taken at 21 K showing dispersive CEF split states. \textbf{b,} Waterfall EDC plot with the extracted dispersion of the CEF split states overlaid on top. \textbf{c,} EDCs taken at the momentum region shown by red arrows in panel \textbf{a,} also highlighted in red in panel \textbf{b,} with their corresponding fits showing individual contributions. A Shirley-type background that has been subtracted prior to the fitting process to account for the inelastically scattered electrons is shown in brown. The binding energy separation between the CEF split states ($\Delta_{1}$, $\Delta_{2}$) at the two different k points are also indicated, highlighting their \textit{k} dependence. In addition to statistical error of one standard deviation from the fitting process, error bars also include variability in the fit results by holding one and/or two peak positions constant in the multi-peak fitting process.}
\label{fig:CEF}
\end{figure*}



\newpage

\section*{References}

\begin{thebibliography}{99}

\bibitem{Schroeder:00} A.\ Schr\"{o}der \textit{et al.}, Onset of antiferromagnetism in heavy-fermion metals, \textit{Nature}, \textbf{407,} 351-355 (2000)

\bibitem{Andres:75} K.\ Andres, J.\ E.\ Graebner, and H.\ R.\ Ott, 4\textit{f}-Virtual-Bound-State Formation in CeAl$_{3}$ at Low Temperatures, \emph{Phys. Rev. Lett.}, \textbf{35,} 1779 (1975)

\bibitem{Mydosh:11} J.\ M.\ Mydosh and P.\ M.\ Oppeneer, Hidden Order, superconductivity, and magnetism : The unsolved case of URu$_{2}$Si$_{2}$, \emph{Rev. Mod. Phys.} \textbf{83,} 1301 (2011)

\bibitem{Curro:05} N.\ J.\  Curro \textit{et al.,} Unconventional superconductivity in PuCoGa$_{5}$, \emph{Nature}, \textbf{434,} 622-625 (2005)

\bibitem{Si:01} Q.\ Si, S.\ Rabello, K.\ Ingersent, and J.\ L.\ Smith, Locally critical quantum phase transitions in strongly correlated metals, \emph{Nature}, 413, 804 (2001)

\bibitem{Gegenwart:08} P.\ Gegenwart, F.\ Steglich, and F. and Q.\ Si, Quantum criticality in heavy-fermion metals, \emph{Nat. Phys.} \textbf{4,} 186?197 (2008)

\bibitem{Si:10} Q.\ Si and F. Steglich, Heavy Fermions and Quantum Phase Transitions, \emph{Science}, \textbf{329,} 5996 (2010)

\bibitem{Varma:76} C.\ M.\ Varma, Mixed-valence compounds. \emph{Rev. Mod. Phys.} \textbf{2,} 244-248 (1976).

\bibitem{Lawrence:80}J.\ M.\ Lawrence, P.\ S.\ Riseborough, and R.\ D.\ Parks, Valence fluctuation phenomena, \emph{Rep. Prog. Phys.} \textbf{44,} 1 (1980)

\bibitem{Parks:77} R.\ D.\ Parks (ed.), Valence Instabilities and narrow band phenomena, Plenum Press, New York (1977)

\bibitem{Tjeng:93} L.\ H.\ Tjeng \emph{et al.}, Temperature dependence of the Kondo Resonance in YbAl$_{3}$, \emph{Phys. Rev. Lett.}, \textbf{71,} 1419 (1993)

\bibitem{Moreschini:07} L.\ Moreschini \emph{et al.}, Comparison of bulk-sensitive spectroscopic probes of Yb valence in Kondo systems, \emph{Phys. Rev. B}, \textbf{75,} 035113 (2007)

\bibitem{Kummer:11} K.\ Kummer \emph{et al.}, Intermediate valence in Yb compounds probed by 4 \textit{f} photoemission and
resonant inelastic x-ray scattering, \emph{Phys. Rev. B}, \textbf{84,} 245114 (2011) 

\bibitem{Suga:05} S.\ Suga \emph{et al.}, Kondo Lattice Effects of YbAl$_{3}$ Suggested by Temperature Dependence of High-Accuracy High-Energy Photoelectron Spectroscopy, \emph{J. Phys. Soc. Jpn.}, \textbf{74,} 2880-2884 (2005)

\bibitem{Bauer:04} E.\ Bauer \emph{et al.}, Anderson lattice behavior in Yb$_{1-x}$Lu$_{x}$Al$_{3}$, \emph{Phys. Rev. B}, \textbf{69,} 125102 (2004)

\bibitem{Lawrence:94} J.\ M.\ Lawrence, G.\ H.\ Kwei, P.\ C.\ Canfield, J.\ G.\ DeWitt, A.\ C.\ Lawson, LIII x-ray absorption in Yb compounds: Temperature dependence of the valence, \textit{Phys. Rev. B}, \textbf{49,} 1627 (1994)

\bibitem{Cornelius:02} A.\ L.\ Cornelius \emph{et al.}, Two Energy Scales and Slow Crossover in YbAl$_{3}$, \emph{Phys. Rev. Lett.}, \textbf{88,} 11 (2002)

\bibitem{Ebihara:03} T.\ Ebihara \emph{et al.}, Dependence of the Effective Masses in YbAl$_{3}$ on Magnetic Field and Disorder, \emph{Phys. Rev. Lett.}, \textbf{90,} 16 (2003)

\bibitem{Wahl:11} P.\ Wahl \emph{et al.}, Local spectroscopy of the Kondo lattice YbAl$_{3}$: Seeing beyond the surface with scanning tunneling microscopy and spectroscopy, \emph{Phys. Rev. B}, \textbf{84,} 245131 (2011)

\bibitem{Chatterjee:16} S.\ Chatterjee \emph{et al.}, Epitaxial growth and electronic properties of mixed valence YbAl$_{3}$ thin films, \emph{J. Appl. Phys.}, \textbf{120,} 035105 (2016)

\bibitem{Supplementary} see Supplementary Information

\bibitem{Ebihara:00} T.\ Ebihara \emph{et al.}, Heavy Fermions in YbAl$_{3}$ Studied by the de Haas-van Alphen Effect, \emph{J. Phys. Soc. Jpn.}, \textbf{69,} 895-899 (2000)

\bibitem{Choi:12} H.\ C.\ Choi \emph{et al.}, Temperature-Dependent Fermi Surface Evolution in Heavy Fermion CeIrIn$_{5}$, \emph{Phys. Rev. Lett.}, \textbf{108,} 016402 (2012)

\bibitem{Kummer:15} K.\ Kummer \emph{et al.}, Temperature-Independent Fermi Surface in the Kondo Lattice YbRh$_{2}$Si$_{2}$, \emph{Phys. Rev. X}, \textbf{5,} 011028 (2015)

\bibitem{Yang_1:08} Y.\ -F.\ Yang and D.\ Pines, Universal Behavior in Heavy-Electron Materials, \emph{Phys. Rev. Lett.}, \textbf{100,} 096404 (2008)

\bibitem{Yang_2:08} Y.\ F.\ Yang \emph{et al.}, Scaling the Kondo Lattice, \emph{Nature}, \textbf{454,} 611-613 (2008)

\bibitem{Yang:11} Y.\ -F.\ Yang and D.\ Pines, Emergent states in heavy electron materials, \emph{Proc. Natl. Acad. Sci. USA}, \textbf{109,} E3060-E3066 (2012)

\bibitem{Burdin:09} S.\ Burdin and V.\ Zlatic, Multiple temperature scales of the periodic Anderson model: Slave boson approach, \emph{Phys. Rev. B}, \textbf{79,} 115139 (2009)

\bibitem{Burdin:00} S.\ Burdin,  A.\ Georges, and D.\ R.\ Grempel, Coherence Scale of the Kondo Lattice, \emph{Phys. Rev. Lett.}, \textbf{85,} 1048 (2000)

\bibitem{Lea:62} K.\ R.\ Lea, M.\ J.\ M.\ Leask, and W.\ P.\ Wolf, The raising of angular momentum degeneracy of \textit{f}-electron terms by cubic crystlal fields. \emph{J. Phys. Chem. Solids.}, \textbf{23,} 1381-1405 (1962)

\bibitem{Murani:94} A.\ Murani, Paramagnetic scattering from the valence-fluctuation compound YbAl$_{3}$, \emph{Phys. Rev. B}, \textbf{50,} 9882 (1994)

\bibitem{Osborn:99} R.\ Osborn \emph {et al.}, Inelastic neutron scattering study of the spin dynamics of Yb$_{1-x}$Lu$_{x}$Al$_{3}$, \emph{J. Appl. Phys.}, 85, 5344 (1999)

\bibitem{Christianson:06} A.\ D.\ Christianson \emph{et al.}, Localized Excitation in the Hybridization Gap in YbAl$_{3}$, \emph{Phys. Rev. Lett.}, \textbf{96,} 0117206 (2006)

\bibitem{Shirley:72} D.\ A.\ Shirley, High-Resolution X-Ray Photoemission Spectrum of the Valence Bands of Gold, \emph{Phys. Rev. B}, \textbf{5,} 4709 (1972)

\bibitem{Wien2k} P.\ Blaha, K.\ Schwartz, G.\ Madsen, D.\ Kvasnicka, and J.\ Luitz, Wien2k, An Augmented Plane Wave Plus Local Orbitals Program for Calculating Crystal Properties, (Karlheinz Schwarz, Techn. Universit\"{a}t Wien, Austria, 2001, ISBN 3-9501031-1-2)

\bibitem{Perdew:96} J.\ P.\ Perdew, K.\ Burke, and M.\ Ernzerhof, Generalized Gradient Approximation Made Simple, \emph{Phys. Rev. Lett.}, \textbf{77,} 3865 (1996) 

\bibitem{Anisimov:93} V.\ I.\ Anisimov, I.\ V.\ Solovyev, M.\ A.\ Korotin, M.\ T.\ Czyzyk, and G.\ A.\ Sawatzky, Density-functional theory and NiO photoemission spectra, \emph{Phys. Rev. B}, \textbf{48,} 16929 (1993)

\bibitem{Finkelstein:16} K.\ D.\ Finkelstein, C.\ J.\ Pollock, A.\ Lyndacker, T.\ Krawcyk, and J.\ Conrad, Dual-Array Valence Emission Spectrometer (DAVES): a New Approach for Hard X-ray Photon-in Photon-out Spectroscopies, \textit{AIP Conf. Proc.}, \textbf{1741,} 030009 (2016)

\bibitem{Zwicknagl:93} G.\ Zwicknagl, Quasiparticles in heavy fermion systems, \textit{Physica Scripta}, \textbf{T49,} 34-41 (1993)

\bibitem{Kumar:08} R.\ S.\ Kumar \emph{et al.}, Pressure-induced valence change in YbAl$_{3}$: A combined high-pressure inelastic x-ray scattering and theoretical investigation, \textit{Phys. Rev. B}, \textbf{78,} 075117 (2008)

\bibitem{Glatzel:05} P.\ Glatzel, and U.\ Bergmann, High resolution 1s core hole X-ray spectroscopy in 3d transition metal complexes: electronic and structural information, \textit{Coord. Chem. Rev.}, \textbf{249,} 65-95 (2005)

\bibitem{Yamaoka:13} H.\ Yamaoka \emph{et al.}, Valence transitions in the heavy-fermion compound YbCuAl as a function of temperature and pressure, \textit{Phys. Rev. B}, \textbf{87,} 205120 (2013)

\bibitem{Jiang:15} W.\ B.\ \emph{et al.}, Crossover from a heavy fermion to intermediate valence state in non-centrosymmetric Yb$_{2}$Ni$_{12}$(P,As)$_{7}$, \textit{Sci. Rep.}, \textbf{5,} 17608 (2015)

\end{thebibliography}
\end{document}